\documentclass[aps,prd,twocolumn,superscriptaddress,nofootinbib,nobibnotes]{revtex4-1}

\usepackage{amssymb}
\usepackage{graphicx}
\usepackage{amsmath,bm}
\usepackage{hyperref}
\usepackage{subfigure}
\usepackage{multirow}
\usepackage{setspace}
\usepackage{verbatim}
\usepackage{float}
\usepackage{color}
\usepackage{ulem}
\usepackage[utf8]{inputenc}
\usepackage[table,xcdraw]{xcolor}
\usepackage{makecell}

\newcommand{\slath}[1]{{/\!\!\!#1}}

\newcommand{\eg}{e.g.,\ }

\newcommand{\beq}{\begin{equation} \begin{aligned}}
\newcommand{\eeq}{\end{aligned} \end{equation}}

\begin{document}

\title{Berry phase in axion physics, SM global structure, and generalized symmetries}

\author{Qing-Hong Cao}
\email{qinghongcao@pku.edu.cn}
\affiliation{School of Physics, Peking University, Beijing 100871, China}
\affiliation{School of Physics, Zhengzhou University, Zhengzhou 450001, China}
\affiliation{Center for High Energy Physics, Peking University, Beijing 100871, China}

\author{Shuailiang Ge}
\email{shuailiangge@kaist.ac.kr}
\affiliation{Department of Physics, Korea Advanced Institute of Science and Technology (KAIST), Daejeon 34141, South Korea}
\affiliation{Center for High Energy Physics, Peking University, Beijing 100871, China}
\affiliation{School of Physics, Peking University, Beijing 100871, China}

\author{Yandong Liu}
\email{ydliu@bnu.edu.cn}
\affiliation{Key Laboratory of Beam Technology of Ministry of Education, School of Physics and Astronomy, Beijing Normal University, Beijing, 100875, China}
\affiliation{Institute of Radiation Technology, Beijing Academy of Science and Technology, Beijing 100875, China}

\author{Jun-Chen Wang}
\email{junchenwang@stu.pku.edu.cn}
\affiliation{School of Physics, Peking University, Beijing 100871, China}

\begin{abstract}
We investigate the Berry phase arising from axion-photon and axion-fermion interactions. The effective Hamiltonians in both systems share the same form, enabling a unified description of the Berry phase and providing a novel perspective on axion experiments. We conceptually propose a new photon-ring experiment for axion detection. Furthermore, we demonstrate that measuring the axion-induced Berry phase offers a unique method for probing the global structure of the Standard Model gauge group and axion-related generalized symmetries.
\end{abstract}

\maketitle

\noindent \textit{\textbf{Introduction}}--
The axion, originally proposed to resolve the strong CP problem~\cite{peccei1977cp, peccei1977constraints, weinberg1978new, wilczek1978problem, kim1979weak, shifman1980can, dine1981simple, Zhitnitsky:1980tq},
has stimulated considerable theoretical and experimental studies, focusing on axion-fermion~\cite{Budker:2013hfa, Berlin:2023ubt, Brandenstein:2022eif, Graham:2020kai, Agrawal:2022wjm, Dror:2022xpi, JEDI:2022hxa} and axion-photon~\cite{Sikivie:1983ip, Sikivie:1985yu, Wuensch:1989sa, ADMX:2018ogs, HAYSTAC:2018rwy, McAllister:2017lkb, Alesini:2019ajt, Kahn:2016aff, Liu:2018icu, CAST:2004gzq, Raffelt:1987im, Galanti:2022yxn, Cao:2023kdu, Fujita:2020ecn, Fedderke:2019ajk, Obata:2018vvr, Nagano:2019rbw, Oshima:2023csb} interactions,
\begin{align}
    \label{eq:fermion Lagrangian}
    \mathcal{L}_{aff}&=-\frac{1}{2}\frac{g_f}{f_a}\partial_{\mu} a\overline{f}\gamma^{\mu}\gamma^{5}f,\\
    \label{eq:photon Lagrangian}
    \mathcal{L}_{a\gamma\gamma}&=\frac{1}{4}\frac{g_{\gamma}}{f_a}aF^{\mu \nu}\tilde{F}_{\mu \nu},
\end{align}
where $a$ is the axion field, $f$ denotes fermions, $F^{\mu\nu}$ is the photon field strength, $f_a$ is the axion decay constant, and $g_{f,\gamma}$ are Wilson coefficients.
Recently, a few theoretical studies refined the quantization of axion couplings with the gauge fields and revealed their connection to the {\it global} structure of the Standard Model (SM) gauge group and generalized symmetries~\cite{Cordova:2023her, Choi:2023pdp, Reece:2023iqn, Agrawal:2023sbp,Alonso:2024pmq,Li:2024nuo,Koren:2024xof}. 
Another important concept that uniquely depends on the {\it topological/global} properties of quantum systems is the so-called Berry phase~\cite{Berry1984quantal, Aharonov:1987gg, Samuel:1988zz, moore1990non, Xiao:2009rm, Baggio:2017aww}, which has been used in the search for axion-photon interactions~\cite{Hoseini:2019woh, Lambiase:2022rto}.
That inspires us to link the Berry phase in axion physics with the SM global structure and generalized symmetries.

Two intrinsic properties of axion are crucial for generating the Berry phase:
1) \textit{Pseudo-scalar nature and time-reversal violation}: The axion is a pseudo-scalar particle with intrinsic parity $-1$ and time-reversal $-1$. A non-degenerate system with symmetry described by an anti-linear operator (e.g., time reversal) does not exhibit a Berry phase~\cite{Baggio:2017aww}; therefore, the absence of time-reversal invariance is essential for the background axion field to induce a Berry phase.
2)\textit{Periodic nature and topological non-triviality}: As a (pseudo) Nambu-Goldstone boson, the axion possesses a periodicity $a \sim a + 2\pi f_a$, which implies that the axion field resides on a space $\mathbb{S}^1$ rather than $\mathbb{R}^1$.
The nontrivial topology enables the axion to induce a Berry phase absent in general scenarios.
In the Letter, we demonstrate that the Berry phase allows one to determine axion-photon coupling without suppression by the decay constant $f_a$ and provides valuable information for SM global structure and axion-related generalized symmetries.
As a by-product, we reinterpret existing axion detection experiments—such as those based on photon birefringence~\cite{Obata:2018vvr, Nagano:2019rbw, Oshima:2023csb} and storage-ring setups~\cite{Graham:2020kai, Agrawal:2022wjm, Dror:2022xpi, JEDI:2022hxa}—through the lens of the Berry phase. Furthermore, inspired by proton-ring configurations and the unified Hamiltonian of the axion-photon and axion-fermion systems, we propose a novel photon-ring experiment that could serve as a new avenue for axion detection.

\noindent \textit{\textbf{Effective Hamiltonians}}-- The Berry phase is evident in the Hamiltonian framework. We first present a unified framework to describe axion-photon and axion-fermion systems and show that Berry phases arise ubiquitously in these two systems.
For the axion-fermion system in Eq.~\eqref{eq:fermion Lagrangian}, the non-relativistic Hamiltonian is~\cite{Sikivie:2020zpn}:
\begin{equation}
\label{eq:fermion-hamilton}
H_{aff}=\frac{g_f}{2 f_a}\left( \nabla a +\partial_t a\frac{\boldsymbol{p}}{m_f}\right) \cdot \boldsymbol{\sigma},
\end{equation}
where $\boldsymbol{p}$, $m_f$, and $\boldsymbol{\sigma}$ represent fermion momentum, mass, and Pauli matrices. Heuristically, we express the axion-photon system analogously and obtain the Hamiltonian:
\begin{equation}
\label{eq:photon Hamiltonian}
H_{a\gamma\gamma}=\frac{g_{\gamma}}{2f_a}\dot{a}(t)\frac{1}{|\boldsymbol{k}|}\boldsymbol{k}\cdot \boldsymbol{S},\qquad \dot{a}(t)\equiv \frac{\mathrm{d}a}{\mathrm{d}t},
\end{equation}
where $\boldsymbol{k}$ is the photon momentum, $\boldsymbol{S}$ is the spin operator for spin-1 systems; see Appendix~\ref{appendix:deriving_photon_H} for details. Therefore, the effective Hamiltonians for the axion-fermion system in Eq.~\eqref{eq:fermion-hamilton} and the axion-photon system in Eq.~\eqref{eq:photon Hamiltonian} exhibit a unified form:
\begin{equation}
	\label{eq:General Hamiltonian}
	H(t) = \boldsymbol{V}(t) \cdot \boldsymbol{j},
\end{equation}
where $\boldsymbol{V}(t) \equiv (V_x, V_y, V_z)$ is a time-dependent vector, and $\boldsymbol{j} \equiv (j_x, j_y, j_z)$ is the spin operator. 
The Berry phase generation can be categorized into two scenarios: changes in the magnitude or the direction of $\boldsymbol{V}(t)$ over time.

\noindent \textit{\textbf{Scenario I: time-varying magnitude}}--
This scenario is typically applicable in cases where photons or fermions are immersed in or traverse through a varying axion background. 
Without loss of generality, we use the axion-photon system as a typical example, assuming that the photon travels in the axion background in the $+z$ direction. Then, Eq.~\eqref{eq:photon Hamiltonian} becomes
\begin{equation}
    \label{eq:photon_Hamiltonian_z_1}
    H_{a\gamma\gamma}=\frac{g_{\gamma}}{2f_a}\dot{a}(t) \begin{pmatrix}
    0 & -i & 0\\
    i & 0 &0 \\
    0&0&0
    \end{pmatrix}.
\end{equation}
We focus only on the upper-left $2 \times 2$ submatrix, as its third component is trivial.
For a non-adiabatic but periodic system,
i.e., $H(t) = H(t+T)$ with a period of $T$, the time evolution operator $U(t)$ is expressed as $U(t) = Z(t)e^{iMt}$ using the operator decomposition method~\cite{moore1990non}, where $M$ is time-independent and Hermitian, and $Z(t)$ satisfies $Z(t+T) = Z(t)$~\cite{moore1990floquet}. For a eigenstate $\left| \phi \right\rangle$ of $M$ with the eigenvalue $\xi$, $U(T) \left| \phi \right\rangle = e^{i\xi T} \left| \phi \right\rangle$, and the phase $\xi T$ acquired over one period comprises both the dynamical and Berry phases~\cite{moore1991calculation}:
\begin{align}
    \alpha_{\rm dyn} &= -i \int_0^{T} \left\langle \phi \right| U^{\dagger}(t) \frac{d}{dt} U(t) \left| \phi \right\rangle dt,     \label{eq:general decomposition dynamical} \\
    \alpha_{\rm Berry} &= i \int_{0}^{T} \left\langle \phi \right| Z^{\dagger}(t) \frac{d}{dt} Z(t) \left| \phi \right\rangle dt. 
     \label{eq:general decomposition Berry}
\end{align}
As shown in Appendix~\ref{appendix:non-adiabatic_Berry_phase}, the time evolution operator $U(t)$ for the axion-photon interaction Hamiltonian $H_{a\gamma\gamma}$ is
\begin{equation}
    \label{eq:new_T_form}
     U(t)
    =\begin{pmatrix}
    \cos \tilde{\beta} & -\sin \tilde{\beta} \\
    \sin \tilde{\beta} &  \cos \tilde{\beta} 
    \end{pmatrix}
    \exp \left[{-i\frac{g_{\gamma}}{2f_a}A\begin{pmatrix}
    0 & -i\\
    i &  0
    \end{pmatrix}t}\right],
\end{equation}
under the assumption that the system is periodic in time, i.e., $\dot{a}(t) = \dot{a}(t+T)$. Here, we define $\tilde{\beta} \equiv \tilde{a}(t) g_{\gamma}/(2f_a)$, where the periodic function $\tilde{a}(t)$ and constant $A$ are determined by integrating the periodic function $\dot{a}(t)$:
\begin{equation}
    \int_0^{t} \dot{a}(t^\prime) \,\mathrm{d} t^\prime = \tilde{a}(t) + A t.
\end{equation}
Matching $U(t)$ to $Z(t)e^{iMt}$, we calculate the Berry phase from Eq.~\eqref{eq:general decomposition Berry}, yielding
\begin{equation}
    \label{eq:Berry_axion_photon}
    \alpha_{\rm Berry}=\pm\frac{g_{\gamma}}{2f_a}\left[\tilde{a}(T)-\tilde{a}(0)\right],
\end{equation}
where the positive and negative sign corresponds to the right-handed circularly polarized state $\left|\psi_R\right> = (1,i)^T/\sqrt{2}$ and the left-handed circularly polarized state $\left|\psi_L\right> = (1,-i)^T/\sqrt{2}$, respectively.
Consider a linearly polarized photon, a superposition of $\left|\psi_R\right>$ and $\left|\psi_L\right>$, traveling in the axion background. The photon polarization vector undergoes a rotation by an angle of magnitude \(|\alpha_{\rm Berry}|\), a phenomenon known as axion-induced photon birefringence~\cite{Harari:1992ea, Carroll:1989vb}, which can be detected in terrestrial experiments~\cite{Obata:2018vvr, Nagano:2019rbw, Oshima:2023csb}. 

The non-trivial topology of the axion field gives rise to a condition $a(t+T) = a(t) + 2\pi N_w f_a$~\cite{Marsh:2015xka}, which induces $\tilde{a}(t+T) = \tilde{a}(t) + 2\pi N_w f_a$ with $N_w$ being an integer representing the winding number around $\mathbb{S}^1$. As a result, the Berry phase in Eq.~\eqref{eq:Berry_axion_photon} takes the form
\begin{equation}
    \label{eq:Berry_axion_photon_final}
    \alpha_{\rm Berry}=\pm N_w\pi g_{\gamma},
\end{equation}
arising as the system completes a closed loop $N_w$ times.
For an axion domain wall with wall number $N_{\rm DW} = 1$, the axion field experiences a discrete shift of $\Delta a / f_a = 2\pi$ across the wall. When a linearly polarized photon traverses such an axion wall, its polarization direction undergoes a rotation with a magnitude given by $|\alpha_{\rm Berry}| = \pi g_{\gamma}$ \footnote{
For the QCD axion that also couples to gluon, an axion wall is accompanied by a pion wall~\cite{Huang:1985tt,Blasi:2024xvj,Agrawal:2019lkr}, which will additionally induce an $-\alpha_{\rm EM}/2$ polarization rotation.
}. For $N_{\rm DW} \neq 1$, although it is not a closed-loop across a wall, the polarization rotation still exists, which is the non-cyclic Berry phase~\cite{Samuel:1988zz}. 
Another approach to realize the condition $a(t+T) = a(t) + 2\pi N_w f_a$ relies on the oscillation of the axion field, provided that its amplitude $a_0$ exceeds $2\pi f_a$. However, in the context of axion dark matter, the energy density of the axion background is approximately $\rho_{\rm DM} \sim 0.3~{\rm GeV}/{\rm cm}^3$, and the corresponding amplitude, given by $a_0 \simeq \sqrt{2\rho_{\rm DM}/m_a^2}$ (where $m_a$ is the axion mass), is typically smaller than $f_a$. This suggests that the oscillation axion dark matter is unlikely to generate berry phase.

The same argument also goes for fermion with the Hamiltonian in Eq.~\eqref{eq:photon Hamiltonian} replaced by Eq.~\eqref{eq:fermion-hamilton}. In the rest frame of fermion, the second term in Eq.~\eqref{eq:fermion-hamilton} can be dropped off, and in the first term $\nabla = \boldsymbol{v}^{-1} \partial_t$ where $\boldsymbol{v}$ is the velocity of axion background. 
Analogous to Eq.~\eqref{eq:Berry_axion_photon_final}, 
the Berry phase that the fermion's spin acquires is
\begin{equation} \label{eq:Berry_axion_ferimion_final_two}
    \alpha_{\rm Berry}=2\pi j_z N_w  g_f v^{-1},\ j_z=\pm \frac{1}{2},
\end{equation}
where the direction and magnitude of $\boldsymbol{v}$ are defined as $+z$ and $v$, respectively.
Considering a fermion as a superposition of states with left-handed ($j_z=-1/2$) and right-handed helicities ($j_z = +1/2$), passing through an axion wall, its spin will be rotated by $|\alpha_{\rm Berry}| = \pi g_f v^{-1}$ in magnitude. A similar result was given in Ref.~\cite{Pospelov:2012mt}, which discussed the wall's effect on spins. We have shown that this is exactly a Berry phase.

One final remark in this section concerns the basis-independence of the Berry phase
~\footnote{
We thank the anonymous referee for raising this question.
}.
The derivation above is carried out in the original fermion–photon basis defined in Eqs.~\eqref{eq:fermion Lagrangian}–\eqref{eq:photon Lagrangian}.
A field redefinition constructed by the chiral rotation can be used to remove the axion–fermion (or axion–photon) couplings appearing in these expressions.
In the rotated basis, the fermion acquires a complex mass term, which reproduces exactly the same Berry phase, as explicitly demonstrated in Appendix~\ref{appendix:basis_independence_Berry_phase}.

\begin{figure}
	\includegraphics[width=0.40\textwidth]{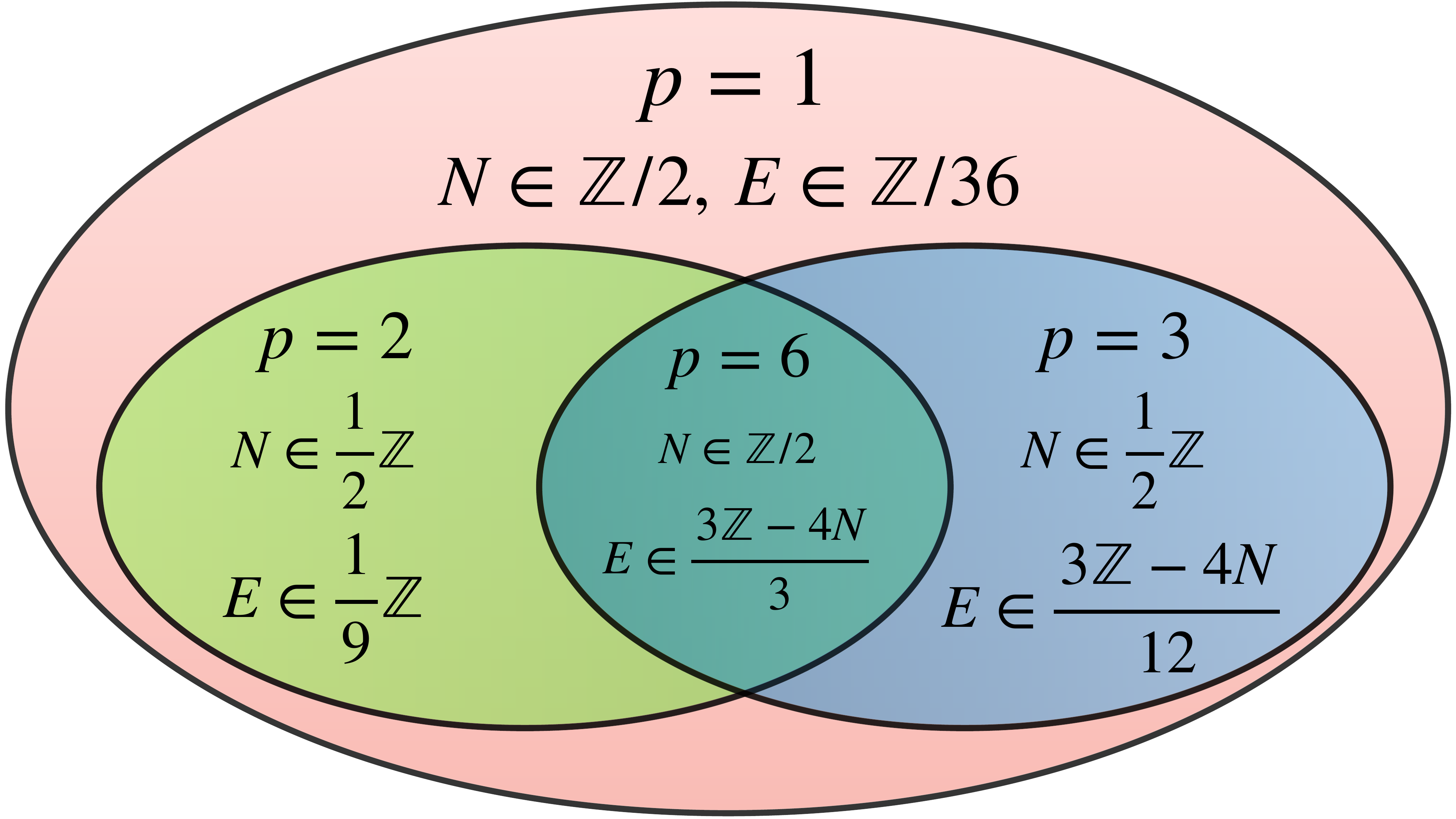}
    \includegraphics[width=0.40\textwidth]{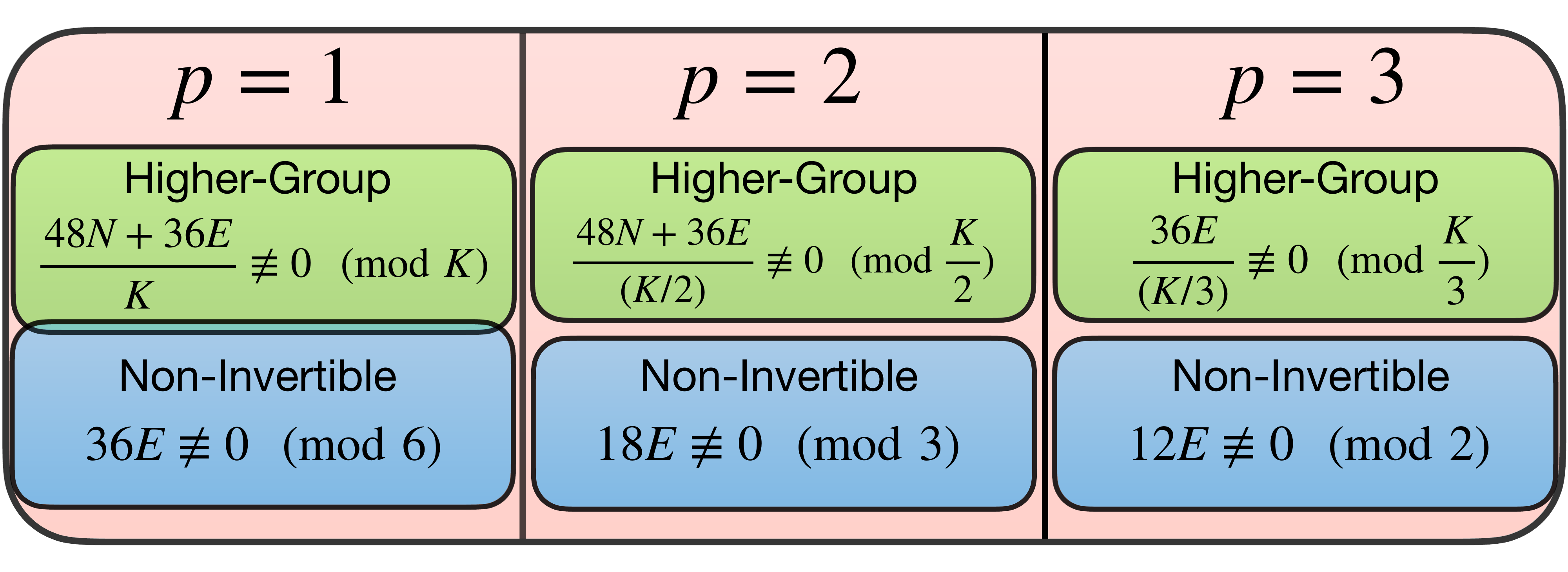}
	\caption{
	Constraints on the quantized couplings $E$ and $N$ for different possible SM global structures $p = 1,2,3,6$ (up) and for realizations of higher-group symmetry and non-invertible symmetry (down). $K \equiv \mathrm{gcd}(6,36E)$ denotes the greatest common factor of $6$ and $36 E$. The constraints are adapted from~\cite{Cordova:2023her, Choi:2023pdp, Reece:2023iqn, Agrawal:2023sbp}. The axion-related higher-group and non-invertible symmetries mentioned here vanish for $p=6$.
	}
	\label{fig:constraint}
\end{figure}

~\\
\noindent \textit{\textbf{Implications for SM global structure and generalized symmetries}}--
In conventional experiments of axion detections, the experimental observables are typically proportional to $g_{\gamma}/f_a$. The signal strength of axion events is suppressed by $f_a$, and it is challenging to disentangle $g_{\gamma}$ from $f_a$.
The Berry phase in Eq.~\eqref{eq:Berry_axion_photon_final} depends only on $g_{\gamma}$ and provides a unique method to measure $g_{\gamma}$. 
The key reason lies in the fact that at the core of a domain wall, the axion field satisfies $a/f_a \sim \pi/N_{\rm DW}$, ensuring that the Berry phase generated by particles crossing an axion wall (or more generally, a closed-loop structure) depends solely on the coupling $g_{\gamma}$. Consequently, this Berry phase serves as a powerful tool for probing non-perturbative aspects of axion physics, including the global structure of the Standard Model (SM) gauge group.
Theoretically, $g_{\gamma}$ is predicted as~\cite{GrillidiCortona:2015jxo}
\begin{equation}
	\label{eq:coefficient}
	g_{\gamma}=\frac{\alpha_{\rm EM}}{\pi}\left(E-1.92N\right),
\end{equation}
$E$ and $N\equiv N_{\rm DW}/2$ originate from the axion-photon and axion-gluon couplings, respectively.
Regardless of many UV models, couplings $E$ and $N$ are quantized due to the axion periodicity $a\sim a +2\pi f_a$~\cite{Cordova:2023her, Choi:2023pdp, Reece:2023iqn, Agrawal:2023sbp}. Furthermore, this quantization differs for different global structures of the SM gauge group, $SU(3)\times SU(2) \times U(1)/Z_p$, with $p=1,2,3,{\rm or}~6$~\cite{Tong:2017oea, Hucks:1990nw, ORaifeartaigh:1986agb}. We have summarized the requirements for $E$ and $N$ in Fig.~\ref{fig:constraint} up-panel based on Refs.~\cite{Cordova:2023her, Choi:2023pdp, Reece:2023iqn, Agrawal:2023sbp}. Therefore, by measuring the Berry phase generated when a photon crosses a closed-loop structure of axion field, one can determine the value of $g_{\gamma}$, which is crucial for inferring the values of $E$ and $N$ and further revealing the SM global structure.

Additional symmetry structures will further constrain the quantization of $E$ and $N$. Axion physics is rich in generalized symmetries, including the higher-group symmetry arising from the mix of axion winding 2-form symmetry and electric 1-form symmetry, and the non-invertible 1-form symmetry as the axion-modified SM center 1-form symmetry consistent with axion periodicity. Constraints on $E$ and $N$ from generalized symmetries are summarized in Fig.~\ref{fig:constraint} down-panel based on Ref.~\cite{Choi:2023pdp}. 
Again, measuring the axion-induced Berry phase for photons provides an opportunity to probe the underlying generalized symmetry structures in axion physics.

Finally, we comment on possible experimental avenues for measuring the Berry phase, providing a few illustrative examples (though this list is not exhaustive). One potential scenario involves an axion string network populating the Universe, which could induce a rotation in the polarization of cosmic microwave background (CMB) photons~\cite{Agrawal:2019lkr, Jain:2021shf}. A similar effect could arise if the Universe is instead populated by axion domain walls~\cite{Takahashi:2020tqv}. Additionally, the passage of axion domain walls through the Earth has been proposed as a possible detection avenue~\cite{Pospelov:2012mt}. Other intriguing possibilities include an axion string piercing a black hole, which can induce the rotation of polarized photons orbiting the black hole due to the axion-photon coupling~\cite{Gussmann:2021mjj}. Furthermore, a large axion field value, $a/f_a \sim \pi$, could accumulate around a neutron star~\cite{Zhang:2021mks}, providing another promising environment for testing axion-induced Berry phases.

One cannot simultaneously  determine both $E$ and $N$ from a single value of $g_{\gamma}$. In the cases discussed in Refs.~\cite{Agrawal:2019lkr, Jain:2021shf, Takahashi:2020tqv}, if the domain wall number $N_{\rm DW}$ is known, then together with the measured photon Berry phase, which determines $g_{\gamma}$, one can subsequently infer the value of $E$. Then, using the relations listed in Fig.~\ref{fig:constraint}, one can infer the information of the global structure. 
Suppose now the value of $N$ is known. For example, the choice $N_{\rm DW} = 1$ (i.e., $N=1/2$) is natural in the sense that it avoids the domain wall problem. 
According to Fig.~\ref{fig:constraint}, the parameter range of $E$ for the cases $p= 2, 3, 6$ is entirely contained within that of $p= 1$. Consequently, a measurement of the Berry phase alone cannot exclude the possibility of $p= 1$.
	Furthermore, for $p= 2, 3, 6$, Fig.~\ref{fig:constraint} shows that when $E$ lies in the interval allowed by $p= 6$, all four cases ($p= 1, 2, 3, 6$) overlap, and thus no specific global structure can be ruled out. However, if $E$ falls outside this overlapping region, it can be determined whether the global structure is $p= 1\ \text{or}\ 2$, $p= 1\ \text{or}\  3$, or uniquely $p= 1$.
	
	Next, we discuss the experimental precision required to distinguish different SM global structures. The intervals between adjacent $E$ values for $p=1,2,3,6$ are respectively $\Delta E = \frac{1}{36}, \frac{1}{9}, \frac{1}{4}, 1$.
	Combining Eqs.~\eqref{eq:Berry_axion_photon_final} and \eqref{eq:coefficient}, we find that the measurement precision of the Berry phase must exceed $\alpha_{\rm EM}\Delta E$ (we take the winding number $N_w = 1$), which are
	$2.0\times10^{-4}$,\ $8.0\times10^{-4}$,\ $2\times10^{-3}$, and $7\times10^{-3}$ for $p=1,2,3,6$, respectively. 
    For example, to uniquely determine $p=1$, the minimal non-overlapping difference $\Delta E_{\rm min}=1/36$ implies that the required measurement precision should be better than $2.0 \times 10^{-4}$.

~\\
\noindent \textit{\textbf{Scenario II: time-varying direction}}--
We now consider the second scenario that  $\boldsymbol{V}(t)$ in Eq.~\eqref{eq:General Hamiltonian} changes in direction while keeping a constant amplitude. For simplicity, we assume $\boldsymbol{V}(t)$ rotates uniformly, $\boldsymbol{V}= V (\sin \theta \cos \omega t,\sin \theta \sin \omega t,\cos \theta)$.  $\theta$ is the angle between $\boldsymbol{V}$ and $+z$ direction. $\omega$ is the angular frequency (period $T=2\pi/\omega$). 
Following the derivation in Appendix~\ref{appendix:non-adiabatic_Berry_phase_scenarioII}, we obtain the Berry phase
\begin{align}
	\label{eq:spin rotate Berry}
	\alpha_{\rm Berry}&=-2\pi j_z(1-\cos \overline{\theta}),\\
    \label{eq:new cos angle}
	\cos \overline{\theta}=&\frac{V\cos \theta-\omega}{\sqrt{V^2+\omega^2-2V\omega \cos \theta}}.
\end{align}
$j_z$ takes the values $ -j, \ -j+1,\ ...,\ j-1,\ j$, with $j$ being the spin quantum number. 
Eq.~\eqref{eq:spin rotate Berry} is a general expression without assuming an adiabatic rotation. Note it recovers the adiabatic result in the limit of $\omega\ll V$.

For a charged particle, the rotation of $\boldsymbol{V}$ can be realized by applying electric and magnetic fields, $\boldsymbol{E}$ and $\boldsymbol{B}$. For $\boldsymbol{B}$ and $\boldsymbol{E}\times \boldsymbol{v}$ in the direction $+z$, the particle velocity is $\boldsymbol{v}(t)=v(\sin\omega t,\cos\omega t, 0)$ with
$\omega=-q\left(B+E/v\right)/(\gamma m_f)$. Here
$q$, $m_f$, and $\gamma \equiv 1/\sqrt{1-v^2}$ are the particle's electric charge, mass, and Lorentz factor, respectively. Using the axion-fermion system as an example, $\boldsymbol{V}$ in Eq.~\eqref{eq:General Hamiltonian} is then
\beq\label{eq:fermion Vz}
    &V_x=\frac{g_f}{f_a}(\partial_t a) v \sin\omega t
    ,~~~
    V_y=\frac{g_f}{f_a}(\partial_t a) v \cos\omega t, \\
    &V_z=-\frac{\tilde{g}q}{2m_f}B-\frac{\tilde{g}q}{2m_f}vE
     +\frac{(\gamma-1)}{2} \omega.
\eeq
$\tilde{g}$ is the Lande factor.
The last term in $V_z$ accounts for the Thomas precession effect.
We have omitted the spatial variation of the axion field, which makes sense in the axion dark matter background because of the low dark matter velocity $v_{\rm DM}\sim 10^{-3}\ll 1$.

Substituting $V_z$ into Eq.~\eqref{eq:new cos angle} and Eq.~\eqref{eq:spin rotate Berry} yields
\begin{equation}
    |\alpha_{\rm Berry}| \simeq \pi j_z \frac{V_x^2 + V_y^2}{(V_z - \omega)^2} \sim \mathcal{O}\left(\left(\frac{g_f}{f_a}\right)^2\right),
\end{equation}
indicating that the Berry phase is significantly suppressed. This suppression arises because two dominant effects—the particle rotation $\omega$ and the electromagnetic contribution $V_z$—overshadow the axion signal.

However, this situation changes dramatically if these two effects cancel each other out, which occurs when $\omega = V_z$. This resonance condition is expressed as
\begin{equation}
\label{eq:resonance condition}
    GB + vE\left(G - \frac{1}{\gamma^2 - 1}\right) = 0,
\end{equation}
where $G \equiv (\tilde{g} - 2)/2$ is the anomalous magnetic moment. Under the resonance condition in Eq.~\eqref{eq:resonance condition}, the angle $\overline{\theta}$ reaches $\pi/2$, leading to the maximal Berry phase in Eq.~\eqref{eq:spin rotate Berry}:
\begin{equation}
    \label{eq:resonance Berry phase}
    \alpha_{\rm Berry} = -2\pi j_z.
\end{equation}

If $B=0$, Eq.~\eqref{eq:resonance condition} is just the magic momentum condition in proton-ring experiments~\cite{Graham:2020kai, Agrawal:2022wjm, Dror:2022xpi, JEDI:2022hxa}. These experiments prepare a proton beam rotating in a ring as a superposition of two cyclic initial states, which will acquire opposite quantum phases from axion dark matter, generating a spin precession. Our discussion above provides a fresh look at these storage-ring experiments. The resonance is met when the Berry phase is maximized. Unfortunately, the axion field does not enter the expression of the maximal Berry phase in Eq.~\eqref{eq:resonance Berry phase}, but instead, the dynamical phase. Following Eq.~\eqref{eq:general decomposition dynamical}, we get
\begin{align} 
\label{eq:fermion total phase value}
&\alpha_{\rm dyn}  
=8.15\times 10^{-7}\ \text{rad}\ \times  \\ \nonumber
&\left(\frac{g_f/f_a}{5\times 10^{-10}\ \text{GeV}^{-1}}\right)\left(\frac{\sqrt{\rho_{\rm DM}}\sin(m_a t_0 +\varphi)}{\sqrt{0.3 \ \text{GeV}\cdot \text{cm}^{-3}}}\right)v\left(\frac{t_{\rm exp}}{1 \text{s}}\right),
\end{align}
where $t_{\rm exp}$ represents the experimental duration time.
The approximation $\partial_t a\simeq \sqrt{2\rho_{\rm DM}}\sin(m_a t_0+\varphi)$ has been used, where $t_0$ denotes the time at which the experiment is performed and $\varphi$ is the initial phase of the axion field. This approximation is valid provided that the axion oscillation period $T_a \equiv 2\pi / m_a$ is much longer than $t_{\rm exp}$.
Note that in the resonance condition, the effect is only suppressed linearly by $g_f/f_a$ rather than quadratically, as expected.

In addition to spin precession~\cite{Graham:2020kai}, the phase effect can be directly measured by interference, \eg electron double-slit experiment. The phase is measured as $\alpha_{\rm dyn} = 2\pi \Delta l d/(\lambda D)$ with $d$ the distance between two slits, $D$ the distance between the baffle and image plane, $\lambda$ the electron wavelength, and $\Delta l$ the fringe shift. For a typical experimental setup $d \sim 1\ \mu$m, $\lambda \sim 1$ pm, $D \sim 1$ mm~\cite{harada2018interference}, to reach the level of $\alpha_{\rm dyn}\sim 10^{-7}$~rad, the resolution of fringe shift should be $\Delta l \sim 10^{-8}$ nm. This tiny value is challenging for current techniques~\cite{harada2018interference}.

\begin{figure}
	\includegraphics[width=0.3\textwidth]{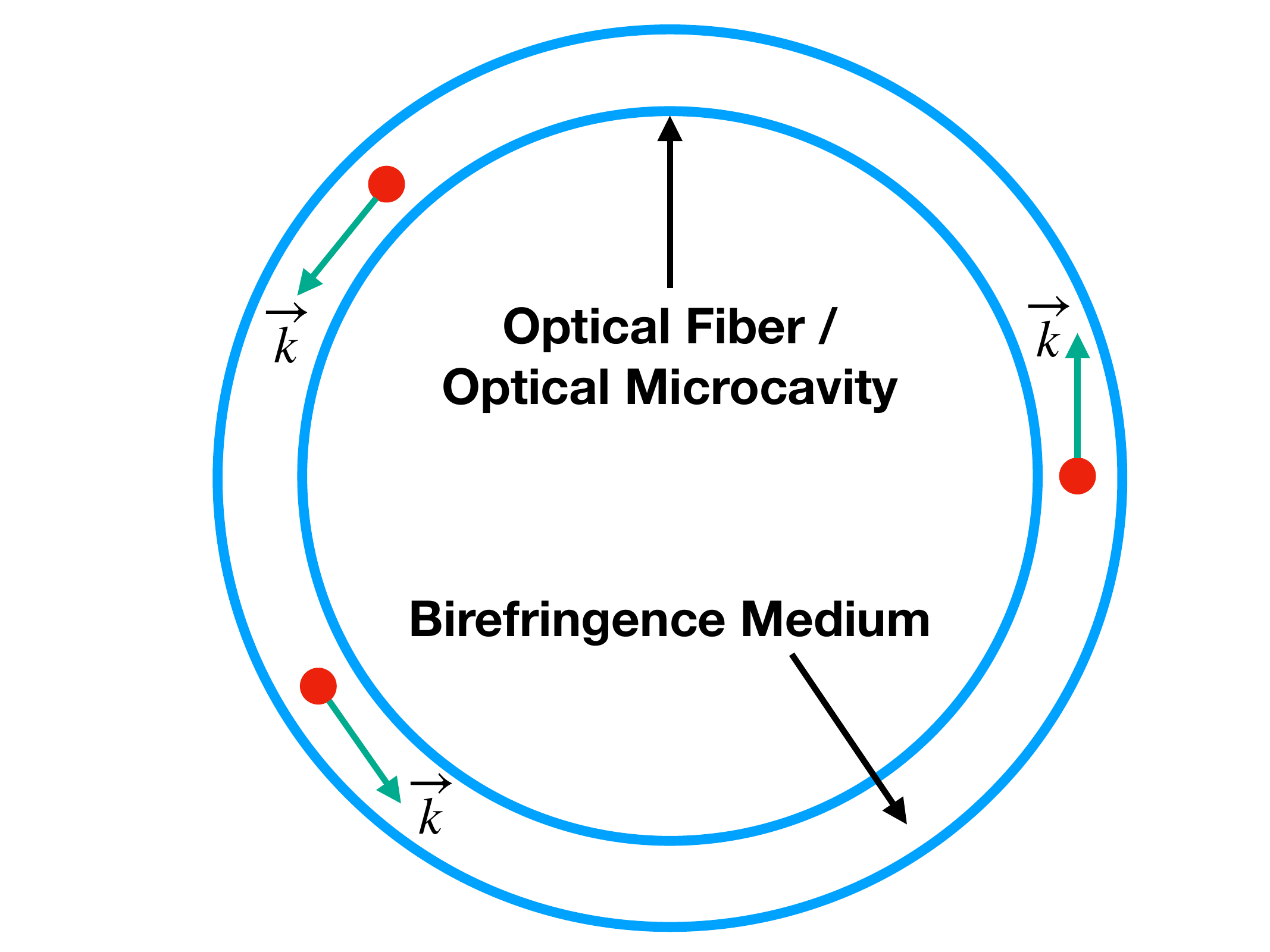}
	\caption{Sketch of the photon-ring experiment. The circular motion of photons can be realized by the optical fiber (or optical microcavity), while the resonance condition is by the birefringence medium. 
 }
	\label{fig:sketch}
\end{figure}

Inspired by the same form of the effective Hamiltonians of axion-fermion and axion-photon systems, we propose a ring-type experiment for photons. 
This experiment is realized by winding an optical fiber into a circular loop and injecting a laser beam linearly polarized perpendicular to the fiber plane (defined as the $x$–$y$ plane). The main setup is sketched in Fig.~\ref{fig:sketch}. Photons propagates circularly within the fiber with momentum $\mathbf{k}(t)=(n\omega \cos (\Omega t), n\omega \sin (\Omega t), 0)$. $\Omega = 1/(nR)$ represents the photon angular velocity. $\omega$, $n$, and $R$ denote the photon energy, optical fiber refractive index, and fiber loop radius, respectively.
In addition, we introduce a birefringent material into the optical fiber with $\chi$ characterizing the birefringence. Then, we have the dielectric permittivity tensor:
    \begin{align}
    \label{eq:special_medium}
    \varepsilon=\begin{pmatrix}
    n^2 & i\chi & 0\\
    -i\chi & n^2 & 0\\
    0 & 0 & n^2
    \end{pmatrix}=n^2 I-\chi S^{z},
    \end{align}
    where $I$ is the identity matrix.
The effective Hamiltonian describing this system is
\begin{equation}
	\label{eq:photon-ring-Hamiltonian}
H(t)=\frac{g_{\gamma}\dot{a}}{2n^2 \omega f_a}\mathbf{k}(t)\cdot \mathbf{S} + \frac{\omega \chi}{2n^2} S^z,
\end{equation}
which is derived in Appendix~\ref{appendix:photon_ring_derivative}. It is a generalization of Eq.~\eqref{eq:photon Hamiltonian} with the medium effects incorporated.
Furthermore, we obtain the dynamical phase that is accumulated by photon with respect to the initial $z$ direction (see Appendix~\ref{appendix:photon_ring_derivative} for more details),
 \begin{align}
 	\label{eq:phase_full}
    \alpha_{\rm dyn}=\frac{t_{\rm exp}\left[\frac{ \rho_{\rm DM} g_{\gamma}^2 \sin^2(m_a t_0 +\varphi)}{2 n^2 f_a^2} -\frac{\omega \chi}{2 n^2}\left(\Omega-\frac{\omega \chi}{2 n^2}\right)\right]}{\sqrt{\frac{\rho_{\rm DM} g_{\gamma}^2 \sin^2(m_a t_0 +\varphi)}{2 n^2 f_a^2}+\left(\Omega-\frac{\omega \chi}{2 n^2}\right)^2}}.
    \end{align}
$t_{\rm exp}$ represents the total duration of the experiment.
The reason that we have introduced a birefringent material is to cancel the phase accumulated from $\Omega$ itself. So the resonance condition can be identified as $\Omega = \omega \chi/(2 n^2).$
Under this resonance condition, the phase is explicitly induced by the axion dark matter background
\begin{equation}
\label{eq:pre_photon_ring}
	\alpha_{\rm dyn} = \frac{g_{\gamma} \sqrt{\rho_{\rm DM}}\sin(m_a t_0 +\varphi)}{\sqrt{2}n f_a}t_{\rm exp}.
\end{equation}
Notably, Eq.~\eqref{eq:pre_photon_ring} closely resembles Eq.~\eqref{eq:fermion total phase value}, highlighting a significant analogy between photon and fermion systems. In practice, this dynamical phase manifests as a rotation of the photon polarization angle, measurable to high precision~\cite{Rowe:2016xwq}, thus presenting a more experimentally feasible approach compared to the fermion-based measurements.
Even though other experiments also measure this photon polarization rotation~\cite{Obata:2018vvr, Nagano:2019rbw, Oshima:2023csb},  there is a crucial difference. In their case, the effect is induced essentially by a varying axion field. In contrast, in our photon-ring case, the variation of photon motion direction plays the role based on scenario II. The circular motion of photons can be realized using the material of optical fiber or optical microcavity. We also need the birefringence material to enforce the resonance condition, as in the case of proton rings.
More details will be presented in a separate paper~\cite{Upcoming_paper_photon_ring}.

The assumption that $\partial_t a$ can be treated as constant requires the experimental time $t_{\rm exp}$ to be much shorter than the axion oscillation period, namely $t_{\rm exp} \ll T_a$. 
With typical experimental parameters, namely a minimal detectable dynamical phase $\alpha_{\rm dyn}\simeq 10^{-9}$~\cite{Rowe:2016xwq}, a fiber refractive index $n\simeq 1.5$, and a local dark-matter density $\rho_{\rm DM}=0.3$ GeV/cm$^{3}$, Eq.~\eqref{eq:pre_photon_ring} implies that probing parameter space beyond existing constraints on the coupling $g_{\gamma} / f_a \simeq 10^{-12}$ GeV$^{-1}$ for ultralight axions~\cite{AxionLimits} requires a minimum experimental time of $t_{\rm exp}^0 \simeq 0.9$ s. 
    To safely treat $\partial_t a$ as constant during the experiment, we conservatively impose $T_a > 10\ t_{\rm exp}^{0} = 9$ s, which translates into an upper bound on the axion mass, $ m_a < 4.5\times 10^{-16}\ \mathrm{eV}$.
    We emphasize that although the phase $\varphi$ cannot be determined in a single measurement, it can be probed experimentally by performing repeated measurements at different times (corresponding to different $t_0$ values). Since the observable signal is proportional to $\partial_t a$, repeated experiments would allow one to track its sinusoidal time dependence. This is beneficial for extracting axion signals in data analysis. A fit to this temporal modulation would also determine the axion oscillation frequency (and hence the mass $m_a$).

As a concluding remark for this section, we comment on the feasibility and the principal challenges of the proposed photon-ring experiment. First, several aspects of current optical technology suggest that such an experiment is in principle viable. Polarization-maintaining optical fibers can preserve linear polarization over long distances and strongly suppress unwanted birefringence originating from temperature gradients or mechanical stress~\cite{noda2003polarization,guan2005stress,chu1984analytical}. This ensures that any observed polarization rotation predominantly reflects the axion-induced signal rather than Standard Model birefringence or material imperfections. Modern polarization-maintaining fibers have therefore become suitable platforms for precision phase measurements.

In addition, the optical medium characterized by the tensor in Eq.~\eqref{eq:special_medium} behaves as a controllable birefringent material. Systems such as crystalline media or stressed fibers naturally support distinct refractive indices for the two orthogonal polarization modes~\cite{chen2018designing, evans1949birefringent,jung2025design}. The birefringence parameter $2\chi$ can be continuously tuned by external pressure, strain, or temperature control~\cite{calvani2002polarization,payne1982development}. Experimental studies have demonstrated that $\chi$ can be stabilized and adjusted at the level of $10^{-9}$–$10^{-5}$~\cite{segura2015low}, which comfortably covers the parameter region relevant for the photon-ring configuration and enables enhancement of the dynamical phase.

Furthermore, photon attenuation along long optical paths can be compensated using solid-state or doped-fiber optical amplifiers~\cite{frede2015catch,aziza2021intelligent}. Such amplifiers restore the photon intensity while leaving the polarization state largely unaffected, making it possible to extend the photon propagation time without significantly degrading coherence. Together with modern low-loss fibers, this allows effective optical lengths sufficient to accumulate a measurable phase shift~\cite{lines1984search,koike2011progress,brambilla2004ultra,ding2019recent}.

Despite these favorable features, several experimental challenges remain. A central difficulty is achieving long coherent propagation times in the optical ring so that the dynamical phase can accumulate to an observable level. From Eq.~\eqref{eq:pre_photon_ring}, the phase scales linearly with the propagation time $t_{\rm exp}$ (or the effective optical length $L_{\rm eff}=t_{\rm exp}/n$). Although optical amplifiers and low-loss fibers can extend $L_{\rm eff}$, maintaining polarization coherence over such distances requires careful characterization and suppression of noise, dispersion, and other decoherence mechanisms. Determining the maximal achievable coherent propagation time therefore represents one of the key practical challenges of the setup.

Another major challenge is satisfying the resonance condition with sufficient accuracy, as this condition suppresses Standard Model contributions while maximizing the axion-induced dynamical phase. In practice, mechanical vibrations, temperature drifts, and slow environmental fluctuations can shift the resonance away from its ideal value. Achieving stable operation thus requires active stabilization, such as precise temperature control, mechanical isolation, and laser-frequency locking, to ensure that relevant optical parameters remain within the narrow tolerance required by the resonance. Assessing the performance of such stabilization techniques is essential for determining the realistic sensitivity of the photon-ring experiment.

~\\
\noindent \textit{\textbf{Conclusion}}--
Due to axion being a pseudoscalar and Nambu-Goldstone boson, the Berry phase naturally arises in axion physics. We present a unified description of the Berry phase in axion-photon and axion-fermion systems, as they share the same effective Hamiltonian form in the low-energy limit.
We study the Berry phase in two scenarios: one in which photons and fermions experience a closed loop of the axion field and the other in which the motion of photons and fermions in the axion background forms a closed loop in the momentum space. 

A nontrivial Berry phase can be derived from axion periodicity. Detecting this Berry phase enables a direct measurement of the dimensionless axion coupling $g_{\gamma}$ without the suppression from $f_a$, which is a portal for probing the SM global structure and axion-related generalized symmetries. 
Additionally, our study of the Berry phase offers new insights into axion experiments, such as photon-birefringence and storage-ring experiments.
Inspired by the unified description of axion-photon and axion-fermion systems, we conceptually proposed a photon-ring experiment, the details of which will be left in our future work~\cite{Upcoming_paper_photon_ring}.

~\\

\noindent \textbf{Acknowledgments:}~We thank Kebin Shi and Yunfeng Xiao for helpful discussions on the feasibility of the photon-ring experiment. The work of Q.-H.C. is partly supported by the National Science Foundation of China under Grant No. 12235001 and by Fundamental and Interdisciplinary Disciplines Breakthrough Plan of the Ministry of Education of China. The work of S.G. is partly supported by the National Research Foundation of Korea (NRF) under Grant No. RS-2024-00405629. The work of Y.L. is partly supported by the National Science Foundation of China under Grant No. 12175016, 12075257 and the National Key R$\&$D Program of China under Grant No. 2023YFA1607104.

\bibliographystyle{apsrev4-1}
\bibliography{ref}

\appendix
\section{Derivation of the axion-photon effective Hamiltonian}\label{appendix:deriving_photon_H}
We derive the effective Hamiltonian governing axion-photon interactions by starting with the standard photon kinetic term, \(-\frac{1}{4}F^{\mu\nu}F_{\mu\nu}\), and the axion-photon coupling term, \(\frac{g_{\gamma}}{4f_a}\,a\,F^{\mu\nu}\tilde{F}_{\mu\nu}\), where \(a\) is the axion field, \(f_a\) is the axion decay constant, and \(g_{\gamma}\) is the coupling constant. Combining these terms modifies Maxwell's equations in the presence of an axion field, as derived in Refs.~\cite{Sikivie:1983ip, Sikivie:1984yz}:
\begin{align}
    \label{eq:axion Maxwell 1}
    \nabla \cdot \left( \boldsymbol{E} - \frac{g_{\gamma}}{f_a}\,a\,\boldsymbol{B} \right) &= 0, \\
    \label{eq:axion Maxwell 2}
    \nabla \times \left( \boldsymbol{B} + \frac{g_{\gamma}}{f_a}\,a\,\boldsymbol{E} \right) &= \partial_t \left( \boldsymbol{E} - \frac{g_{\gamma}}{f_a}\,a\,\boldsymbol{B} \right), \\
    \label{eq:axion Maxwell 3}
    \nabla \cdot \boldsymbol{B} &= 0, \\
    \label{eq:axion Maxwell 4}
    \nabla \times \boldsymbol{E} + \partial_t \boldsymbol{B} &= 0.
\end{align}

To proceed, we adopt a plane-wave ansatz for the electromagnetic fields:
\begin{align}
    \label{eq:ansatz E}
    \boldsymbol{E}(t, \boldsymbol{x}) &= \boldsymbol{E}_0(t) \, e^{-i\omega t + i\boldsymbol{k} \cdot \boldsymbol{x}}, \\
    \label{eq:ansatz B}
    \boldsymbol{B}(t, \boldsymbol{x}) &= \boldsymbol{B}_0(t) \, e^{-i\omega t + i\boldsymbol{k} \cdot \boldsymbol{x}},
\end{align}
where \(\boldsymbol{k}\) is the wave vector and \(\omega = |\boldsymbol{k}|\) is the angular frequency.

We assume the geometric-optics regime, in which the axion field \(a\) varies slowly compared to the electromagnetic oscillation scale. Consequently, second derivatives of \(a\) are negligible~\cite{Harari:1992ea, Sikivie:2020zpn}. Taking the time derivative of Eq.~\eqref{eq:axion Maxwell 2} and substituting Eq.~\eqref{eq:axion Maxwell 4}, we obtain:
\begin{equation}
    \label{eq:derivation 1}
    \partial_t^2 \boldsymbol{E} + \frac{g_{\gamma}}{f_a} \, \partial_t a \, \nabla \times \boldsymbol{E} + \nabla (\nabla \cdot \boldsymbol{E}) - \nabla^2 \boldsymbol{E} - \frac{g_{\gamma}}{f_a} \, \nabla a \times \partial_t \boldsymbol{E} \simeq 0.
\end{equation}

Using Eqs.~\eqref{eq:axion Maxwell 1} and \eqref{eq:axion Maxwell 3}, the term \(\nabla (\nabla \cdot \boldsymbol{E})\) can be expressed as:
\begin{align}
    \label{eq:derivation 2}
    \nabla (\nabla \cdot \boldsymbol{E}) &= \nabla \left( \frac{g_{\gamma}}{f_a} \, \nabla a \cdot \boldsymbol{B} \right)  \nonumber \\
    &\simeq \frac{g_{\gamma}}{f_a} \left[ \nabla a \times (\nabla \times \boldsymbol{B}) + (\nabla a \cdot \nabla) \boldsymbol{B} \right].
\end{align}

Substituting Eqs.~\eqref{eq:axion Maxwell 2}, \eqref{eq:axion Maxwell 4}, and the ansatz Eq.~\eqref{eq:ansatz B} into Eq.~\eqref{eq:derivation 2}, and neglecting terms of order \(\mathcal{O}(g_{\gamma}^2 / f_a^2)\), we find:
\begin{equation}
    \label{eq:derivation 3}
    \nabla (\nabla \cdot \boldsymbol{E}) \simeq \frac{g_{\gamma}}{f_a} \, \nabla \times \partial_t \boldsymbol{E} + \frac{g_{\gamma}}{f_a} \, (\nabla a \cdot \nabla) \, \frac{\nabla \times \boldsymbol{E}}{i\omega}.
\end{equation}

Inserting Eq.~\eqref{eq:derivation 3} into Eq.~\eqref{eq:derivation 1} yields:
\begin{equation}
    \label{eq:derivation 4}
    \partial_t^2 \boldsymbol{E} + \omega^2 \boldsymbol{E} + i \, \frac{g_{\gamma}}{f_a} \left( \partial_t a + \frac{\boldsymbol{k} \cdot \nabla a}{|\boldsymbol{k}|} \right) (\boldsymbol{k} \times \boldsymbol{E}) = 0.
\end{equation}

Since \(\partial_t a + \frac{\boldsymbol{k} \cdot \nabla a}{|\boldsymbol{k}|}\) represents the total time derivative along the photon trajectory, we define \(\dot{a}(t) \equiv \frac{d a}{d t}\). Within the plane-wave ansatz Eq.~\eqref{eq:ansatz E}, the second time derivative becomes:
\begin{align}
    \label{eq:derivation 5}
    \partial_t^2 \boldsymbol{E} &= (-\omega^2 \boldsymbol{E}_0 - 2 i \omega \partial_t \boldsymbol{E}_0 + \partial_t^2 \boldsymbol{E}_0) \, e^{-i\omega t + i\boldsymbol{k} \cdot \boldsymbol{x}} \nonumber \\
    &\simeq (-\omega^2 \boldsymbol{E}_0 - 2 i \omega \partial_t \boldsymbol{E}_0) \, e^{-i\omega t + i\boldsymbol{k} \cdot \boldsymbol{x}},
\end{align}
where the WKB approximation (\(\partial_t^2 \boldsymbol{E}_0 \ll \omega \partial_t \boldsymbol{E}_0\)) justifies neglecting the \(\partial_t^2 \boldsymbol{E}_0\) term. Substituting into Eq.~\eqref{eq:derivation 4}, we obtain:
\begin{equation}
    \label{eq:derivation 6}
    \partial_t \boldsymbol{E}_0 = \frac{g_{\gamma}}{2 f_a} \, \dot{a}(t) \, \frac{\boldsymbol{k} \times \boldsymbol{E}_0}{|\boldsymbol{k}|}.
\end{equation}

Expressing Eq.~\eqref{eq:derivation 6} in component form:
\begin{equation}
    \label{eq:derivation 7}
    i \partial_t E_{0,\alpha} = i \frac{g_{\gamma}}{2 f_a} \, \dot{a}(t) \, \frac{1}{|\boldsymbol{k}|} \, \epsilon^{\alpha \beta \delta} k_{\beta} E_{0,\delta},
\end{equation}
where \(\epsilon^{\alpha \beta \delta}\) is the Levi-Civita symbol (\(\alpha, \beta, \delta = 1, 2, 3\)). Defining the spin-1 matrices \((S^{\beta})_{\alpha \delta} = i \epsilon^{\alpha \beta \delta}\), which satisfy:
\[
    [S^{\alpha}, S^{\beta}] = i \epsilon^{\alpha \beta \delta} S^{\delta}, \quad (S^1)^2 + (S^2)^2 + (S^3)^2 = 2,
\]
we treat \(\boldsymbol{E}_0 = (E_{0,1}, E_{0,2}, E_{0,3})^T\) as a quantum state \(|\psi\rangle\). Equation~\eqref{eq:derivation 7} then takes the Schrödinger form:
\begin{equation}
    \label{eq:derivation 8}
    i \partial_t |\psi\rangle = H_{a\gamma\gamma} |\psi\rangle,
\end{equation}
with the effective axion-photon Hamiltonian:
\begin{equation}
    \label{eq:photon Hamiltonian app}
    H_{a\gamma\gamma} = \frac{g_{\gamma}}{2 f_a} \, \dot{a}(t) \, \frac{\boldsymbol{k} \cdot \boldsymbol{S}}{|\boldsymbol{k}|}.
\end{equation}
This matches the Hamiltonian presented in the main text.

\section{Derivation of Non-Adiabatic Berry Phase in Scenario I}
\label{appendix:non-adiabatic_Berry_phase}

Consider a photon propagating along the \(\hat{z}\)-direction in an axion background. The Hamiltonian from Eq.~\eqref{eq:photon Hamiltonian app} simplifies to:
\begin{equation}
    \label{eq:photon_Hamiltonian_z_app}
    H_{a\gamma\gamma} = \frac{g_{\gamma}}{2 f_a} \, \dot{a}(t) \, S^3 = \frac{g_{\gamma}}{2 f_a} \, \dot{a}(t) \begin{pmatrix} 0 & -i & 0 \\ i & 0 & 0 \\ 0 & 0 & 0 \end{pmatrix}.
\end{equation}

Since the third component is decoupled, we focus on the \(2 \times 2\) sub-block:
\begin{equation}
    \label{eq:photon_Hamiltonian_z_1_app}
    H_{a\gamma\gamma} = \frac{g_{\gamma}}{2 f_a} \, \dot{a}(t) \begin{pmatrix} 0 & -i \\ i & 0 \end{pmatrix}.
\end{equation}

The corresponding time-evolution operator is:
\begin{equation}
    \label{eq:photon_Hamiltonian_z_0_app}
    U(t) = \begin{pmatrix} \cos \left( \frac{g_{\gamma}}{2 f_a} \Delta a \right) & -\sin \left( \frac{g_{\gamma}}{2 f_a} \Delta a \right) \\ \sin \left( \frac{g_{\gamma}}{2 f_a} \Delta a \right) & \cos \left( \frac{g_{\gamma}}{2 f_a} \Delta a \right) \end{pmatrix},
\end{equation}
where \(\Delta a(t) = \int_0^t \dot{a}(t') \, dt'\). Assuming \(\dot{a}(t)\) is periodic with period \(T\), we decompose \(\Delta a(t)\) as:
\begin{equation}
    \label{eq:new_form_app}
    \Delta a(t) = \tilde{a}(t) + A t,
\end{equation}
where \(\tilde{a}(t)\) is \(T\)-periodic and \(A\) is a constant determined by \(\dot{a}(t)\). Substituting into Eq.~\eqref{eq:photon_Hamiltonian_z_0_app}, we write \(U(t) = Z(t) e^{i M t}\), with:
\begin{equation}
    \label{eq:ZM_value_app}
    Z(t) = \begin{pmatrix} \cos \tilde{\beta} & -\sin \tilde{\beta} \\ \sin \tilde{\beta} & \cos \tilde{\beta} \end{pmatrix}, \quad M = -\frac{g_{\gamma}}{2 f_a} A \begin{pmatrix} 0 & -i \\ i & 0 \end{pmatrix},
\end{equation}
and \(\tilde{\beta} \equiv \frac{g_{\gamma}}{2 f_a} \tilde{a}(t)\).

The cyclic states, eigenstates of \(M\), are \(\left| v_L \right\rangle = (1, -i)^T / \sqrt{2}\) (left-circular) and \(\left| v_R \right\rangle = (1, i)^T / \sqrt{2}\) (right-circular). The Berry phase is:
\begin{align}
    \label{eq:Berry_axion_photon_app}
    \alpha_{\text{Berry}} &= i \int_0^T \left\langle v_{L/R} \right| Z^{\dagger}(t) \frac{d}{dt} Z(t) \left| v_{L/R} \right\rangle \, dt \nonumber \\
    &= m \frac{g_{\gamma}}{2 f_a} \left[ \tilde{a}(T) - \tilde{a}(0) \right], \quad m = \pm 1.
\end{align}

Since \(a\) is periodic up to \(2\pi N_w f_a\) (with \(N_w\) the winding number on \(\mathbb{S}^1\)), we find:
\begin{equation}
    \label{eq:Berry_axion_photon_final_app}
    \alpha_{\text{Berry}} = \pm N_w \pi g_{\gamma},
\end{equation}
consistent with the main text.

\section{Basis-independence of the Berry phase}
\label{appendix:basis_independence_Berry_phase}
	
In this appendix, we demonstrate explicitly that the axion-induced Berry phases obtained in the main text are independent of the basis choice. Our main point is that the effective nonrelativistic Hamiltonians used in the main text are derived at the classical (tree-level) level, whereas the chiral rotation that removes the axion–fermion Yukawa interaction necessarily invokes the quantum anomaly, a loop effect. Consequently, to maintain physical equivalence between different bases, the loop-induced contributions—particularly the anomaly-generated axion–photon coupling—must be taken into account.

    To make this explicit, we start from a DFSZ-type axion model~\cite{Dine:1981rt,Zhitnitsky:1980tq,Buen-Abad:2021fwq}. The relevant Yukawa interaction is
	\begin{equation}
		\label{eq:Yukawa}
		\mathcal{L}=g_{0} \left( \phi \psi_{L}^{\dagger} \psi_{R}+\phi^{*} \psi_{R}^{\dagger} \psi_{L}\right). 
	\end{equation}
	$\psi_{L/R}$ denotes a left-/right-handed SM fermion, and $\phi$ is a complex scalar charged under the $U(1)_{\rm PQ}$ symmetry. 
    For simplicity, here we consider a toy model with one Dirac fermion.
    After spontaneous breaking of $U(1)_{\rm PQ}$ symmetry. The PQ complex scalar $\phi$ can be written as
	\begin{equation}
		\label{eq:SSB}
		\phi = f_a e^{i\frac{a}{f_a}},
	\end{equation}
	where $a$ is the axion field and $f_a$ is the symmetry breaking scale. Substituting Eq.~\eqref{eq:SSB} into Eq.~\eqref{eq:Yukawa}, and defining the Dirac spinor $\psi \equiv (\psi_{L}, \psi_{R})^{T}$, we obtain
	\begin{equation}
		\label{eq:Yukawa_new}
		\mathcal{L}=g_0 f_a \overline{\psi}e^{i\gamma^{5}\frac{a}{f_a}}\psi.
	\end{equation}
After performing the chiral rotation $\psi \rightarrow e^{-i\gamma^{5}\frac{a(x)}{2f_a}}\psi$, the axion-fermion interaction in \eqref{eq:Yukawa_new} is rotated away, while new interaction terms occur -- namely Eqs.~\eqref{eq:fermion Lagrangian} and \eqref{eq:photon Lagrangian} in the main text. In the main text, we have already demonstrated the presence of the Berry phase in the basis defined by Eqs.~\eqref{eq:fermion Lagrangian} and \eqref{eq:photon Lagrangian}. We now show that the Berry phase is the same in the original basis \eqref{eq:Yukawa_new}.
To do so, we expand the exponential and retain terms up to $\mathcal{O}(a/f_a)$:
	\begin{equation}
		\label{eq:Yukawa_new_new}
			\mathcal{L}=g_0 f_a \overline{\psi}\psi+ig_{0} a \overline{\psi} \gamma^{5} \psi.
	\end{equation}
	The first term contributes to the fermion mass $m_f \equiv g_0 f_a$, as demonstrated in the DFSZ-type models~\cite{Dine:1981rt,Zhitnitsky:1980tq,Buen-Abad:2021fwq}. In what follows, we derive both the photon Berry phase and the fermion Berry phase from the interaction term in Eq.~\eqref{eq:Yukawa_new_new}.

	\begin{figure}[t]
		\centering
		\includegraphics[width=0.49\linewidth]{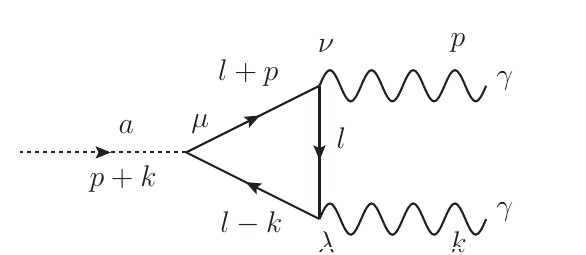}
		\includegraphics[width=0.49\linewidth]{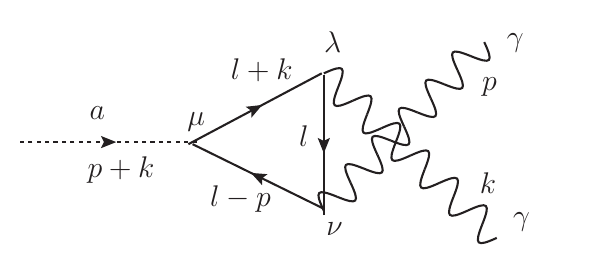}
		\caption{
			One loop diagrams of the axion-photon interaction.
		}
		\label{fig:Process00}
	\end{figure}

	\begin{figure}[t]
		\centering
		\includegraphics[width=0.5\linewidth]{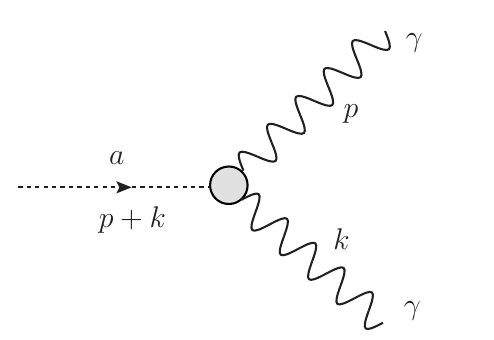}
		\caption{
			Effective interaction vertex between the axion and photon.
		}
		\label{fig:Process01}
	\end{figure}

    The photon-axion interaction in the basis \eqref{eq:Yukawa_new} or \eqref{eq:Yukawa_new_new} involves the one-loop diagrams shown in Fig.~\ref{fig:Process00}. The corresponding amplitude is
	\begin{equation}
		\label{eq:amplitude_g0}
		\mathcal{M} = \epsilon^{*}_{\nu}(p)\epsilon^{*}_{\lambda}(k)\mathcal{M}^{\nu \lambda}(p,k).
	\end{equation}
	$\epsilon^{*}$ represents the photon polarization vector. The expression for $\mathcal{M}^{\nu \lambda}(p,k)$ is
	\begin{align}
		\label{eq:amplitude_new_g0}
		&\mathcal{M}^{\nu \lambda}
		= 2e^2 g_0 \int_{0}^{1} d x \int_{0}^{1} dy \int \frac{d^{4}l}{(2\pi)^4} \\
        &\frac{\text{Tr}\left[\gamma^{5}(\slath{l}-\slath{k}+m_f)\gamma^{\lambda}(\slath{l}+m_f)\gamma^{\nu}(\slath{l}+\slath{p}+m_f)\right]}{\left[(l-xk+yp)^2-(xk-yp)^2+xk^2+yp^2-m_f^2\right]^3} \nonumber \\ 
		 +& \frac{\text{Tr}\left[\gamma^{5}(\slath{l}-\slath{p}+m_f)\gamma^{\nu}(\slath{l}+m_f)\gamma^{\lambda}(\slath{l}+\slath{k}+m_f)\right]}{\left[(l-xp+yk)^2-(xp-yk)^2+xp^2+yk^2-m_f^2\right]^3}. \nonumber
	\end{align}
	We now evaluate the $\gamma$-matrix traces. Using standard identities, we find
	\beq
		\label{eq:tr_g0}
		&\text{Tr}\left[\gamma^{5}(\slath{l}-\slath{k}+m_f)\gamma^{\lambda}(\slath{l}+m_f)\gamma^{\nu}(\slath{l}+\slath{p}+m_f)\right] \\
        & =4i m_f k_{\alpha}p_{\beta}\epsilon^{\alpha \lambda \nu \beta}, \\ 
		&\text{Tr}\left[\gamma^{5}(\slath{l}-\slath{p}+m_f)\gamma^{\nu}(\slath{l}+m_f)\gamma^{\lambda}(\slath{l}+\slath{k}+m_f)\right] \\
        &= 4im_f p_{\alpha}k_{\beta}\epsilon^{\alpha \nu \lambda \beta} = 4im_f k_{\alpha}p_{\beta}\epsilon^{\alpha  \lambda  \nu \beta}.
	\eeq
	Hence,
	\beq	\label{eq:amplitude_new_g0_new}
		&\mathcal{M}^{\nu \lambda}=8ie^2 g_0m_f  k_{\alpha}p_{\beta}\epsilon^{\alpha  \lambda  \nu \beta}\int_{0}^{1} d x \int_{0}^{1} dy \int \frac{d^{4}l}{(2\pi)^4}\\
		&  \frac{1}{\left[(l-xk+yp)^2-(xk-yp)^2+xk^2+yp^2-m_f^2\right]^3} \\  
        + & \frac{1}{\left[(l-xp+yk)^2-(xp-yk)^2+xp^2+yk^2-m_f^2\right]^3} .
	\eeq
	For on-shell photons, we have $p^2=k^2 =0$. Additionally, energy–momentum conservation for $a \to \gamma \gamma$ gives $p\cdot k = m_a^2 / 2$,
    here we take $m_a\ll m_{f}$, like the standard $\pi_0\rightarrow \gamma\gamma$ calculation in~\cite{Schwartz:2014sze}.
    Then, shifting $l \to l + xk -yp$ in the first term of Eq.~\eqref{eq:amplitude_new_g0_new} and $l \to l + xp -yk$ in the second term, we get
	\beq
    \label{eq:amplitude_new_g0_new_new}
		\mathcal{M}^{\nu \lambda}&=16ie^2 g_0m_f  k_{\alpha}p_{\beta}\epsilon^{\alpha  \lambda  \nu \beta}\\
        &
        \times \int_{0}^{1} d x \int_{0}^{1} dy \int \frac{d^{4}l}{(2\pi)^4} \frac{1}{(l^2-m_f^2)^3}.
	\eeq
	Performing a Wick rotation and dimensional regularization, we have
	\beq
		\label{eq:amplitude_new_g0_new_new_new}
		\mathcal{M}^{\nu \lambda}
		&=16ie^2 g_0m_f  k_{\alpha}p_{\beta}\epsilon^{\alpha  \lambda  \nu \beta} \frac{\Gamma\left(3-d/2\right)}{\Gamma(3)(4\pi)^{d/2}} \\
        &\times 
        \int_{0}^{1} d x \int_{0}^{1} dy \left(\frac{1}{m_f^2}\right)^{3-d/2}.
	\eeq
	Taking the limit $d \to 4$, we obtain
	\begin{equation}
		\label{eq:amplitude_final_g0}
		\mathcal{M}^{\nu \lambda} = \frac{ie^2 g_0}{2 \pi^2 m_f}k_{\alpha}p_{\beta}\epsilon^{\alpha  \lambda  \nu \beta}.
	\end{equation}
	Thus, the one-loop diagrams in Fig.~\ref{fig:Process00} match the effective interaction
	\begin{equation}
		\label{eq:effecitve_g0_interaction}
		\mathcal{L_{\rm eff}}=\kappa a F^{\mu \nu}\tilde{F}_{\mu \nu}.
	\end{equation}
    This effective interaction is shown in Fig.~\ref{fig:Process01}, with the amplitude
	\begin{equation}
		\label{eq:amplitude_g0_eff}
		\mathcal{M}_{\rm eff} = \epsilon^{*}_{\nu}(p)\epsilon^{*}_{\lambda}(k)\mathcal{M}^{\nu \lambda}_{\rm eff}(p,k),
	\end{equation}
	where
	\begin{equation}
		\label{eq:amplitude_g0_eff_new}
		\mathcal{M}^{\nu \lambda}_{\rm eff} = - 4i \kappa k_{\alpha} p_{\beta} \epsilon^{\alpha \nu  \lambda \beta} .
	\end{equation}
	Comparing Eq.~\eqref{eq:amplitude_final_g0} with Eq.~\eqref{eq:amplitude_g0_eff_new} fixes
	\begin{equation}
		\kappa = \frac{e^2 g_0}{8 \pi^2 m_f}
	\end{equation} 
	Therefore, we arrive at the effective axion-photon interaction induced by the loop effects, given by:
	\begin{equation}
		\label{eq:effecitve_g0_interaction_new}
		\mathcal{L_{\rm eff}}=\frac{e^2 g_0}{8 \pi^2 m_f} a F^{\mu \nu}\tilde{F}_{\mu \nu},
	\end{equation}
	which is just Eq.~\eqref{eq:photon Lagrangian} in the main text, with $g_{\gamma}= e^2 g_0 f_a / (2 \pi^2 m_f)$. After recovering the axion-photon interaction, it is straightforward to recover the axion-induced photon Berry phase by following the same steps as those outlined in the main text.

We now turn to recover the axion-induced fermion Berry phase in the basis Eq.~\eqref{eq:Yukawa_new_new}. To obtain the axion–fermion scattering amplitude $\mathcal{M}_{aff}$ generated by $ig_0 a \overline{\psi} \gamma^5 \psi$, we make the standard replacement
	\begin{equation}
		\label{eq:spinor_new}
		\psi \to u(p)e^{-ip \cdot x},\ \ u(p)=\left(\sqrt{p\cdot \sigma}\ \xi, \sqrt{p\cdot \overline{\sigma}}\ \xi \right)^T,
	\end{equation}
	with $\sigma = (I, \bm{\sigma})$, $\overline{\sigma} =(I, -\bm{\sigma})$ where $I$ is the identity matrix and $\bm{\sigma}$ are the Pauli matrices. Here $\xi = (1, 0)^{T}$ ($\xi = (0,1)^{T}$) denotes spin up (down) along the $z$ direction. In the non-relativistic limit,
	\beq
		\label{eq:non_relativity}
		&e^{-i p \cdot x}\simeq e^{-i m_f t + i \mathbf{p} \cdot \mathbf{x}}, \\
        &
        u(p) \simeq \sqrt{m_f} \left(\left(I-\frac{\mathbf{p}\cdot \bm{\sigma}}{2m_f} \right) \xi, \left(1+\frac{\mathbf{p}\cdot \bm{\sigma}}{2m_f}\right) \xi \right)^T,
	\eeq
	where $\bm{p}$ is the fermion three-momentum. The amplitude then becomes
	\beq
		\label{eq:non_rela_new}
		\mathcal{M}_{aff} &=ig_0 a e^{-i(p-p')\cdot x}\overline{u}(p') \gamma^{5} u(p) \\ 
		& \simeq 2  ig_0 a e^{i(\bm{p}-\bm{p'})\cdot \bm{x}}m_f \xi^{\dagger}\left(\frac{\bm{p}\cdot \bm{\sigma}}{2m_f} - \frac{\bm{p'}\cdot \bm{\sigma}}{2m_f} \right) \xi \\ 
		&= ig_0 a e^{i\bm{k}\cdot \bm{x}} \xi^{\dagger} (\mathbf{k}\cdot \bm{\sigma}) \xi,
	\eeq
	where $\mathbf{k} \equiv \mathbf{p} - \mathbf{p'}$ is the transferred momentum. Using $\mathbf{k} e^{i\mathbf{k} \cdot \mathbf{x}} = -i(\nabla e^{i\mathbf{k} \cdot \mathbf{x}})$ and integrating by parts (dropping boundary terms), we obtain
	\begin{align}
		\label{eq:amp_aff}
		\mathcal{M}_{aff}
        = -g_0 e^{i\mathbf{k} \cdot \mathbf{x}} (\nabla a) \cdot (\xi^{\dagger} \bm{\sigma} \xi).
 	\end{align}
 	Comparing $\mathcal{M}_{aff}$ with the Born approximation in nonrelativistic quantum mechanics, $\mathcal{M}_{aff}=-m_f(\xi^{\dagger}H_{\rm int} \xi) e^{i\mathbf{k}\cdot \mathbf{x}} / (2 \pi)$, we can identify the effective Hamiltonian $H_{\rm int}$ for the axion–fermion interaction as
 	\begin{equation}
 		\label{eq:Ham}
 		H_{\rm int} = \frac{2\pi g_0}{m_f} (\nabla a) \cdot \bm{\sigma},
 	\end{equation}
 	which corresponds to the effective Hamiltonian $H_{aff} = g_f (\nabla a) \cdot \bm{\sigma} / (2 f_a)$ used in the main text~\footnote{Eq.~\eqref{eq:Ham} reproduces the first term of Eq.~\eqref{eq:fermion-hamilton} in the main text; keeping higher orders in the expansion of Eq.~\eqref{eq:non_relativity} will yield the second term of Eq.~\eqref{eq:fermion-hamilton}.}, 
    with $g_f = 4 \pi g_0 f_a / m_f$. With this effective Hamiltonian recovered, again, it is straightforward to recover the axion-induced fermion Berry phase by repeating the steps outlined in the main text.

\section{Derivation of Non-Adiabatic Berry Phase in Scenario II}
\label{appendix:non-adiabatic_Berry_phase_scenarioII}

Consider a time-dependent Hamiltonian \(H(t) = \boldsymbol{V}(t) \cdot \boldsymbol{j}\), where \(\boldsymbol{V}(t)\) has constant magnitude and rotates uniformly:
\[
    \frac{\boldsymbol{V}(t)}{|\boldsymbol{V}|} = \left( \sin \theta \cos \omega t, \sin \theta \sin \omega t, \cos \theta \right),
\]
with period \(T = 2\pi / \omega\). Following Ref.~\cite{Wang:1990vb}, \(H(t) = e^{-i j_z \omega t} H_0 e^{i j_z \omega t}\), where \(H_0 = H(0)\). The Schrödinger equation \(i \partial_t |\psi(t)\rangle = H(t) |\psi(t)\rangle\) transforms to:
\begin{equation}
    \label{eq:reduced equation}
    i \frac{\partial}{\partial t} |\tilde{\psi}(t)\rangle = H(\omega) |\tilde{\psi}(t)\rangle,
\end{equation}
where \(|\psi(t)\rangle = e^{-i j_z \omega t} |\tilde{\psi}(t)\rangle\) and \(H(\omega) = H_0 - \omega j_z\). 
This corresponds precisely to a coordinate transformation from the original, static coordinate system to a comoving frame that rotates synchronously with the vector $\mathbf{V}(t)$.
Defining \(\overline{\boldsymbol{V}}\) such that:
\[
    H(\omega) = \overline{\boldsymbol{V}} \cdot \boldsymbol{j}, \quad \overline{\boldsymbol{V}} = \overline{V} (\sin \overline{\theta}, 0, \cos \overline{\theta}),
\]
with
\begin{equation}
	\label{eq:V_new}
    \overline{V} = \sqrt{V^2 + \omega^2 - 2 V \omega \cos \theta}, 
\end{equation}
and
\[
\sin \overline{\theta} = \frac{V \sin \theta}{\overline{V}}, \quad \cos \overline{\theta} = \frac{V \cos \theta - \omega}{\overline{V}}.
\]

Thus, Eq.~\eqref{eq:reduced equation} becomes time-independent, yielding:
\begin{equation}
    \label{eq:gs}
    |\psi(t)\rangle = e^{-i j_z \omega t} e^{-i H(\omega) t} |\psi(0)\rangle.
\end{equation}

The time-evolution operator is:
\begin{equation}
    \label{eq:new time evolution operator}
    U(t) = e^{-i j_z \omega t + i j \omega t} e^{-i [H(\omega) t + j \omega t]},
\end{equation}
matching \(U(t) = Z(t) e^{i M t}\), with:
\[
    M = -[H(\omega) + j \omega], \quad Z(t) = e^{i (j - j_z) \omega t}.
\]

The eigenvalues \(\xi_z\) and eigenstates \(|\phi_z\rangle\) of \(M\) are:
\begin{equation}
    \label{eq:Homega eigen-value}
    \xi_z = -j_z \overline{V} - j \omega, \quad |\phi_z\rangle = e^{-i \overline{\theta} j_y} |j_z\rangle,
\end{equation}
where \(j_z |j_z\rangle = j_z |j_z\rangle\) (\(j_z = -j, -j+1, \dots, j\)). The Berry phase is:
\begin{align}
    \label{eq:appendix_spin rotate Berry}
    \alpha_{\text{Berry}} &= i \int_0^{2\pi / \omega} \left\langle \phi_z \right| e^{-i (j - j_z) \omega t} \frac{d}{dt} e^{i (j - j_z) \omega t} \left| \phi_z \right\rangle \, dt \nonumber \\
    &= 2\pi (j_z \cos \overline{\theta} - j) = -2\pi j_z (1 - \cos \overline{\theta}).
\end{align}
Additionally, we have the total phase
\begin{equation}
	\label{eq:tot spin}
	\alpha_{\rm tot} = j_z \overline{V} T.
\end{equation}

\section{Derivation of $\alpha_{\rm tot}$ in the photon ring experiment}\label{appendix:photon_ring_derivative}

Including the axion–photon coupling together with the standard photon kinetic term, we obtain the axion-modified Maxwell equations. Unlike Eqs.~\eqref{eq:axion Maxwell 1}-\eqref{eq:axion Maxwell 4}, we now also include an external charge density $\rho$ and current density $\mathbf{j}$, yielding
\begin{align}
	\label{eq:axion Maxwell 1 new}
	&\nabla\cdot\left(\mathbf{E}-\frac{g_{\gamma}}{f_a}a\mathbf{B}\right)=\rho,\\
	\label{eq:axion Maxwell 2 new}
	&\nabla\times\left(\mathbf{B}+\frac{g_{\gamma}}{f_a}a\mathbf{E}\right)-\partial_t \left(\mathbf{E}-\frac{g_{\gamma}}{f_a}a\mathbf{B}\right)=\mathbf{j},\\
	\label{eq:axion Maxwell 3 new}
	&\nabla\cdot \mathbf{B}=0,\\
	\label{eq:axion Maxwell 4 new}
	&\nabla\times \mathbf{E}+\partial_t \mathbf{B}=0,
\end{align}
For photons propagating in the medium, $\rho$ and $\mathbf{j}$ can be expressed in terms of the electric polarization $\mathbf{P}$ as $\rho=-\nabla \cdot \mathbf{P}$ and $\mathbf{j}=\partial_t \mathbf{P}$. Introducing the electric displacement field $\mathbf{D} \equiv \mathbf{E}+\mathbf{P}$, the equations become
\begin{align}
	\label{eq:axion_Maxwell_1}
	&\nabla\cdot\left(\mathbf{D}-\frac{g_{\gamma}}{f_a}a\mathbf{B}\right)=0,\\
	\label{eq:axion_Maxwell_2}
	&\nabla\times\left(\mathbf{B}+\frac{g_{\gamma}}{f_a}a\mathbf{E}\right)-\partial_t \left(\mathbf{D}-\frac{g_{\gamma}}{f_a}a\mathbf{B}\right)=0,\\
	\label{eq:axion_Maxwell_3}
	&\nabla\cdot \mathbf{B}=0,\\
	\label{eq:axion_Maxwell_4}
	&\nabla\times \mathbf{E}+\partial_t \mathbf{B}=0.
\end{align}
In a general linear medium, $\mathbf{D}=\varepsilon\mathbf{E}$, where the dielectric tensor can be decomposed as
\begin{equation}
	\label{eq:dielectric_constant_matrix}
	\varepsilon=\varepsilon_n I+\kappa,
\end{equation}
with $\varepsilon_n I$ the isotropic part ($\epsilon_n =n^2$) and $\kappa$ the anisotropic part, where $I$ is the identity matrix and $n$ is the refractive index.
As in Scenario II, we neglect the axion momentum and assume $\partial_t a$  varies slowly over the experimental timescale. It can therefore be treated as approximately constant, with $\partial_t a \simeq \sqrt{2\rho_{\rm DM}}\sin(m_a t_0 +\varphi)$, where $t_0$ denotes the time at which the experiment is performed and $\varphi$ is the initial phase of the axion field.
We further adopt the ansatz for the electric field
\begin{equation}
	\label{eq:ansatz E new}
	\mathbf{E}(t,\mathbf{x})=\mathbf{E}^{a}(t)e^{-i\omega t+i\mathbf{k}\cdot \mathbf{x}},
\end{equation}
where $\mathbf{k}$ is the photon wave vector and $\omega=|\mathbf{k}|/n$. Taking a time derivative of Eq.~\eqref{eq:axion_Maxwell_2} gives
\begin{equation}
	\label{eq:derivation 1 new}
	\partial_t^2 \left(\varepsilon \mathbf{E}\right)+ \eta_a \nabla \times \mathbf{E}+\nabla (\nabla \cdot \mathbf{E})-\nabla^2 \mathbf{E}=0,
\end{equation}
where we have defined $\eta_{a}\equiv g_{\gamma}\sqrt{2\rho_{\rm DM}}\sin(m_a t_0+\varphi)/f_a$ for brevity.
Substituting the ansatz Eq.~\eqref{eq:ansatz E new} into Eq.~\eqref{eq:derivation 1 new} yields
\begin{align}
	\label{eq:derivation 2 new}
	&-2i\omega \varepsilon \partial_t \mathbf{E}^{a}+\varepsilon \partial_t^2 \mathbf{E}^{a}- \kappa \omega^2 \mathbf{E}^{a}\\ \nonumber
	&+i\eta_a \mathbf{k}\times \mathbf{E}^{a}-\left(\mathbf{k}\cdot \mathbf{E}^{a}\right)\mathbf{k}=0.
\end{align}
Within the WKB approximation we neglect the second time derivative of $\mathbf{E}^{a}$. Assuming a weak anisotropy, we approximate $-2i\omega \varepsilon \partial_t \mathbf{E}^{a} \simeq -2i\omega \varepsilon_n \partial_t \mathbf{E}^{a}$ and $\mathbf{k}\cdot \mathbf{E}^{a} \simeq 0$. Eq.~\eqref{eq:derivation 2 new} then reduces to
\begin{equation}
	\label{eq:derivation 3 new}
	i\partial_t \mathbf{E}^{a}= -\frac{\omega}{2\varepsilon_n} \kappa \mathbf{E}^{a}+i\frac{\eta_a}{2\varepsilon_n \omega} \mathbf{k}\times \mathbf{E}^{a}.
\end{equation}
Writing $\mathbf{E}^{a}$ and $\kappa$ in their components, we have
\begin{equation}
	\label{eq:derivation 4 new}
	i\partial_t E^{a}_{\alpha}= -\frac{\omega}{2\varepsilon_n}\kappa^{\alpha\beta} E^{a}_{\beta}+i\frac{\eta_a}{2\varepsilon_n \omega} \epsilon^{\alpha \beta \gamma} k_{\beta}E^{a}_{\gamma}.
\end{equation}
As discussed in Appendix~\ref{appendix:deriving_photon_H}, using $i\epsilon^{\alpha \beta \gamma}\equiv (S^{\beta})_{\alpha \gamma}$, the matrices $\mathbf{S}$ are the spin operators with spin quantum number 1. If we regard  $\mathbf{E}^{a}$ as a quantum state, i.e., $(E^{a}_1,E^{a}_2,E^{a}_3)^{T}\equiv \left|\psi\right>$, \eqref{eq:derivation 4 new} becomes
\begin{equation}
	\label{eq:derivation 5 new}
	i\partial_t \left|\psi\right>=-\frac{\omega}{2\varepsilon_n} \kappa \left|\psi\right> +\frac{\eta_a}{2\varepsilon_n \omega}\mathbf{k}\cdot \mathbf{S} \left|\psi\right>,
\end{equation}
which is just the form of the Schr\"odinger equation with the Hamiltonian
\begin{equation}
	\label{eq:photon Hamiltonian new}
	H=-\frac{\omega}{2\varepsilon_n}\kappa+\frac{\eta_a}{2\varepsilon_n \omega}\mathbf{k}\cdot \mathbf{S}.
\end{equation}
For a birefringent medium,
\begin{align}
	\label{eq:special_medium new}
	\kappa=\begin{pmatrix}
		0 & i\chi & 0\\
		-i\chi & 0 & 0\\
		0 & 0 & 0
	\end{pmatrix}=-\chi S^{z},
\end{align}
so that Eq.~\eqref{eq:photon Hamiltonian new} reduces to
\begin{equation}
	\label{eq:photon Hamiltonian new new}
	H=\frac{\omega \chi}{2\varepsilon_n}S^z+\frac{\eta_a}{2\varepsilon_n \omega}\mathbf{k}\cdot \mathbf{S},
\end{equation}
which has the generic form $H(t) = \mathbf{V}(t) \cdot \mathbf{j}$, with $\mathbf{j}=(S^x, S^y, S^z)$ and $\mathbf{V}(t) = (\eta_a/ (2\epsilon_n \omega)) \cdot (k_x, k_y, k_z + \omega^2 \chi / \eta_a)$. For photons circulating in an optical-fiber ring, $(k_x, k_y, k_z) = (n\omega \cos (\Omega t), n\omega \sin (\Omega t), 0)$ with $\Omega = 1/(nR)$ and $R$ being the ring radius. Now using the formulas Eqs.~\eqref{eq:V_new}-\eqref{eq:tot spin} in the Appendix~\ref{appendix:non-adiabatic_Berry_phase_scenarioII}, we obtain
\begin{align}
	\label{eq:final new new}
&\alpha_{\rm tot} = \\ \notag
&t_{\rm exp}\sqrt{\frac{ \rho_{\rm DM} g_{\gamma}^2 \sin^2(m_at_0+\varphi)}{2 n^2 f_a^2}+\left(\Omega-\frac{\omega \chi}{2 n^2}\right)^2}, \\
\label{eq:final_new_new_Berry}
&\alpha_{\rm Berry}= \\ \notag
&\frac{t_{\rm exp}\Omega(\Omega-\omega \chi / (2 n^2))}{\sqrt{\rho_{\rm DM} g_{\gamma}^2 \sin^2(m_at_0+\varphi)/ (2 n^2 f_a^2)+\left(\Omega-\omega \chi / (2 n^2)\right)^2}},\\
\label{eq:final_new_new_dyn}
&\alpha_{\rm dyn}= \\ \notag
&\frac{t_{\rm exp}\left[\frac{ \rho_{\rm DM} g_{\gamma}^2 \sin^2(m_at_0+\varphi)}{2 n^2 f_a^2} -\frac{\omega \chi}{2 n^2}\left(\Omega-\frac{\omega \chi}{2 n^2}\right)\right]}{\sqrt{\frac{ \rho_{\rm DM} g_{\gamma}^2 \sin^2(m_at_0+\varphi)}{2 n^2 f_a^2}+\left(\Omega-\frac{\omega \chi}{2 n^2}\right)^2}}.
\end{align}
Here we have already assumed that the photon circulates around the optical fiber for $N$ loops, with $t_{\rm exp} \equiv 2\pi N / \Omega$ being the experiment duration time. Note that Eq.~\eqref{eq:final_new_new_dyn} is just Eq.~\eqref{eq:phase_full} in the main text.

\end{document}